\documentclass{cup-hpl}

\usepackage[title]{appendix}

\begin{document}

% Enunciations
\newtheorem{theorem}{Theorem}

% Short title and authors in the running heads
\shorttitle{Extremely powerful and frequency-tunable terahertz pulses from a table-top laser-plasma wiggler}                                   
\shortauthor{J. Cai, et al.}

% Title of the document
\title{Extremely powerful and frequency-tunable terahertz pulses from a table-top laser-plasma wiggler}

% Authors with their corresponding address link as [1,2] etc.

\author[1]{Jie Cai}
\author[3]{Yinren Shou}
\author[1]{Yixing Geng}
\author[2]{Liqi Han}
\author[1]{Xinlu Xu}
\author[2]{Shuangchung Wen}
\author[4]{Baifei Shen}

\author[2]{Jinqing Yu \corresp{Jinqing Yu, Hunan Provincial Key Laboratory of High-Energy Scale Physics and Applications, School of Physics and Electronics, Hunan University; Xueqing Yan, State Key Laboratory of Nuclear Physics and Technology, and Key Laboratory of HEDP of the Ministry of Education, CAPT, Peking University.\\
\email{jinqing.yu@hnu.edu.cn (Jinqing Yu); x.yan@pku.edu.cn (Xueqing Yan)}}}
\author[1, 5, 6]{Xueqing Yan}

% Address
\address[1]{State Key Laboratory of Nuclear Physics and Technology, and Key Laboratory of HEDP of the Ministry of Education, CAPT, Peking University}
\address[2]{Hunan Provincial Key Laboratory of High-Energy Scale Physics and Applications, School of Physics and Electronics, Hunan University}
\address[3]{Center for Relativistic Laser Science, Institute for Basic Science, Gwangju}
\address[4]{Shanghai Normal University}
\address[5]{Collaborative Innovation Center of Extreme Optics, Shanxi University}
\address[6]{Guangdong Laser Plasma Institute}

% Abstract
\begin{abstract}
The production of broadband, terawatt terahertz (THz) pulses has been demonstrated by irradiating relativistic lasers on solid targets. However, the generation of extremely powerful, narrow-band, and frequency-tunable THz pulses remains a challenge. Here, we present a novel approach for such THz pulses, in which a plasma wiggler is elaborated by a table-top laser and a near-critical density plasma. In such a wiggler, the laser-accelerated electrons emit THz radiations with a period closely related to the plasma thickness. Theoretical model and numerical simulations predict a THz pulse with a laser-THz energy conversion over 2.0$\%$, an ultra-strong field exceeding 80 GV/m, a divergence angle approximately 20$^\circ$, and a center-frequency tunable from 4.4 to 1.5 THz, can be generated from a laser of 430 mJ. Furthermore, we demonstrate that this method can work across a wide range of laser and plasma parameters, offering potential for future applications with extremely powerful THz pulse.
\end{abstract}

% Keywords
\keywords{Terahertz; Laser; Plasma; Wiggler}

\maketitle

\section{Introduction}\label{sec1}

%In the past decades, the Terahertz (THz) wave \cite{siegel2002}has attracted an explosion of interest due to it's high demand in biology, medicine, material science, communication and military \cite{Zhang201716}. 

%%%%%%%%%%%%%%%%%%%%%%%
%%% my Introduction %%%
%%%%%%%%%%%%%%%%%%%%%%%

Terahertz (THz) waves \cite{siegel2002}, which are significantly less developed and utilized than microwaves and light waves, hence it used to be regarded as 'THz gap'\cite{sirtori2002} in the electromagnetic spectrum due to the limitations of traditional electronic or optical methods in producing terahertz radiation during the early years. Such THz pulses have attracted significant interest for their potential applications in fields such as biology, medicine, material science, optical communication, and the military\cite{RevModPhys.83.543, choi2019terahertz, RN64, koenig2013wireless, kalashnikov2016infrared, Zhang201716}. To generate THz waves, several routine methods such as cascade quantum laser\cite{kazarinov1971possible,faist1994quantum}, optical rectification\cite{yang1971generation,auston1984cherenkov}, photoconductive\cite{auston1984picosecond}, and vacuum electronic\cite{maestrini2005540,crowe1992gaas} have been widely demonstrated. Due to material damage threshold, these methods are incapable of generating extremely powerful THz pulses. Such pulses can be used as a powerfully driven pulse for probing and controlling material properties\cite{kampfrath2013resonant,larue2015thz}, biological macromolecules\cite{choi2019terahertz}, electron beam detection\cite{RN56}, and charged particle acceleration\cite{RN43, RN45}. For instance, the mechanisms of four-wave mixing and photoionization become saturated in under-dense plasma just around the power of $10^{15}$ W/cm$^2$\cite{RN106, RN113, RN112}. As laser intensity continues to improve and nanotechnology develops, terahertz sources based on ultra-intense laser irradiation of dense plasma have started to emerge\cite{hamster1993subpicosecond, gopal2013observation}.

During the past two decades, the quick development of relativistic laser systems, whose peak intensity exceeds $10^{18}$ W/cm$^2$, has opened a new door to obtaining extremely powerful THz pulses\cite{yi2019coherent}. However, the existing terahertz sources based on relativistic lasers lack the ability to tune their spectra, which limits their versatility in applications \cite{vicario2013,baierl2016,zhang2018,balogh2011,yuan2013}. For example, two-color laser filaments can produce THz pulses with field amplitudes above 10 GV/m and a conversion efficiency of 2.36\%\cite{koulouklidis2020observation}, but such pulses have a non-tunable center frequency and a large bandwidth around hundreds of GHz. Another novel method, coherent transition radiation (CTR), occurs when an over-dense electron bunch, produced by irradiating a relativistic laser on a solid target, passes through the vacuum-plasma interface and can emit THz pulses with an energy of 55 mJ and a peak electron field at 4 GV/m\cite{liao2019}. Numerous efforts have been dedicated to achieving tunable strong-field terahertz radiation using various approaches, such as plasma oscillation\cite{pearson2011simulation, miao2017high, kwon2018high} and plasma slab methods\cite{gupta2022coherent}. However, most of these solutions have not been able to reach high field strengths.
C. Miao et al. demonstrated frequency tunable terahertz radiation with weak field strength by manipulating the frequency of plasma waves\cite{miao2017high}. Kyu Been Kwon et al. developed a terahertz source based on plasma dipole oscillation formed by two laser pulses\cite{kwon2018high}.
One of the challenges lies in accessing terahertz frequencies through radiation from plasma waves, as it requires extremely low densities, making it difficult to obtain strong-field terahertz radiation simultaneously.
In a significant breakthrough, G. Liao et al. achieved 0.1-1 THz tunable radiation by transforming the terahertz generation mechanism in laser-irradiated solid targets\cite{liao2020towards}. However, it should be noted that the radiation derived from different mechanisms exhibits distinct characteristics. The center frequency of strong-field terahertz radiations above 1 THz from CTR is also non-adjustable\cite{liao2019review}.

%%%%%%%% Figure 1 %%%%%%%%

\begin{figure}[htb]
    \centering
    \includegraphics[width=\linewidth]{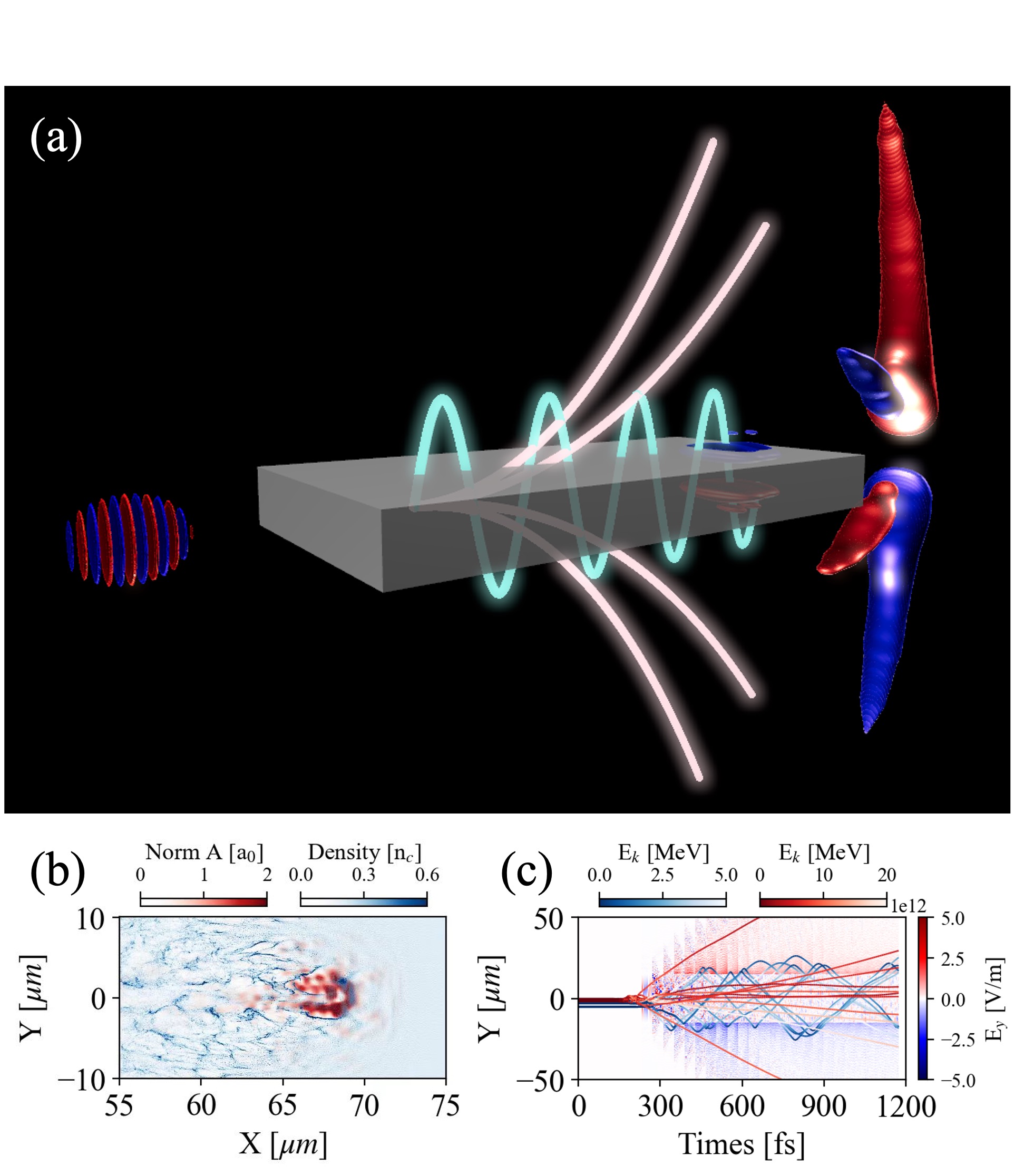}
    \caption{(a) Schematics for the generation of a high-power, collimated, narrow-band and center-frequency tunable THz pulse. An intense femtosecond laser pulse irradiates on the left side of a block-shaped near-critical density plasma. Hot electrons generated by laser ponderomotive force can be separated into two groups: the electrons in group A moving forward leaving the plasma and the electrons in group B reciprocating under the sheath fields $E_s$, here the transverse sheath fields $E_s$ is induced when electrons pass through the plasma transverse interfaces. Under the action of $E_s$, electrons in group B could be pulled back into the plasma and pass through the transverse interface on the other side. Such wiggler-like motions of these electrons can emit the desired THz pulse. (b) The electron accelerating in the plasma, and (c) shows the trajectories of the two group of electrons (blues and reds) in the surface charge separation field.}
    \label{fig1}
\end{figure}

Here, we propose a novel plasma wiggler that uses a femtosecond (fs) laser pulse to produce frequency-tunable and extremely powerful terahertz radiations, improving the flexibility and accuracy of terahertz applications. Fig. 1a draws the concept of the plasma wiggler. A laser pulse is used to irradiate a block-shaped near-critical density plasma, producing hot electrons \cite{wilks1992}. The accelerated electrons move in two different trajectories. One kind of high energy electrons get rid of the potential barrier of the surface electric field and go away from the plasma surface. The other group of electrons are trapped by the barrier acting as a wiggler. Fig. 1b illustrates the laser plasma interaction snapshot at around 273 femtosecond. When this electron beam passes through the transverse interfaces of the plasma, THz radiation \cite{liao2019review} and sheath fields $E_s$ \cite{daido2012} can be induced near the interfaces. Under the influence of $E_s$, the electrons are pulled back into the plasma and cross the transverse interface on the other side, exhibiting reciprocating motion along the plasma. The period of this motion is closely related to the plasma thickness and the size of $E_s$. As a result, the plasma with transverse sheath fields $E_s$ can be used as a wiggler to manipulate the reciprocating motion of the electrons and regulate THz radiations. A theoretical model, which is verified by Particle-in-cell (PIC) simulations, has been developed to describe the physical principles of the tunable THz pulse generation.

%\section{Results}\label{sec2}

Based on the theoretical model and PIC simulations, we find that the center frequency of the THz pulse can be regulated by changing the thickness of the plasma. In simulations using a laser pulse with energy of $\sim430$ mJ, the generated THz pulse has a divergence angle of $\sim20^\circ$, an ultra-strong field strength of over 80 GV/m, a laser-THz conversion efficiency of over 2.0\%, and a center frequency tunable from 4.4 to 1.5 THz by varying the plasma thickness from 20 to 80 $\mu$m. We also demonstrate that this plasma wiggler can work effectively over a wide range of plasma length, thickness, density, and laser intensity. Therefore, this method could overcome significant scientific obstacles in the generation of high-quality THz sources and open up brand-new applications \cite{vicario2013,baierl2016,zhang2018,balogh2011,yuan2013}.

%%%%%%%%%%%%%%%%%%%%%%%%%%%%%%%%%%%%
%%%%%%% End of Introduction %%%%%%%%
%%%%%%%%%%%%%%%%%%%%%%%%%%%%%%%%%%%%

\section{Results}\label{sec2}

\subsection{Theoretical model and simulation setup}\label{subsec1}

When an intense laser pulse irradiates on a plasma as shown in Fig. 1a, hot electrons whose beam length is close to the laser duration can be generated by laser ponderomotive force \cite{wilks1992}. The electron beam transports in the plasma and passes through its transverse interfaces. As a result, transverse sheath fields $E_s$ (more details at supplementary materials) can be generated and the maximum strength \cite{daido2012} on the plasma surfaces can be expressed as
\begin{equation}
  E_s = \sqrt{ \frac {8\pi n_e T_e}{e}},  
\label{equ1}
\end{equation} 
where $e$ is the Euler's number, $n_e$ and $T_e$ are the density and temperature of the hot electrons, respectively. Here, transverse sheath fields $E_s$ higher than $10^{12} $ V/m can be induced near the plasma transverse interfaces \cite{tarkeshian2018} as shown in Figs. 1c. Under the action of $E_s$, electrons whose kinetic energy less than $\varepsilon_e$ could be pulled back into the plasma and pass through the transverse interface on the other side \cite{sentoku2003}, where $\varepsilon_e$ could be calculated by
\begin{equation}
  \varepsilon_e = \int_{0}^{l_e} \frac {E_s}{ \sin{\phi}} dl.
\label{equ2}
\end{equation} 
Here $l$ is the transverse size of $E_s$, $\phi$ is the angle at which the electron enters the transverse sheath fields $E_s$, and $l_e$ is the transverse location where the electron could be pulled back into the plasma. Electron field derived from the simulation was used (more details at supplementary materials) for calculation, and then the relation between $\varepsilon_e$ and $l_e$ could be plotted in Fig. 2 by employing the $E_s$ as shown in Fig. 1c. In the case that the electron perpendicularly penetrates the $E_s$ ($\phi=90^\circ$), the threshold kinetic energy is about 6 MeV. Most of the electrons penetrate the transverse sheath fields with a much smaller $\phi$, and the threshold kinetic energy could be much higher than 6 MeV. Some electrons, which located in the forefront of the electron beam and experienced much weaker $E_s$, could not be pulled back to the plasma even with a much smaller kinetic energy \cite{mora2003}. According to the large difference on the dynamical behavior of the hot electrons as shown in Figs. 1c with red and blue colors, we could separate the hot electrons into two groups: the electrons in group A located in the front of the electron beam marked with red color and the electrons in group B located in the beam rear express as blue color.

%%%%%%%% Figure 2 %%%%%%%%

\begin{figure}[htb]
    \centering
    \includegraphics[width=\linewidth]{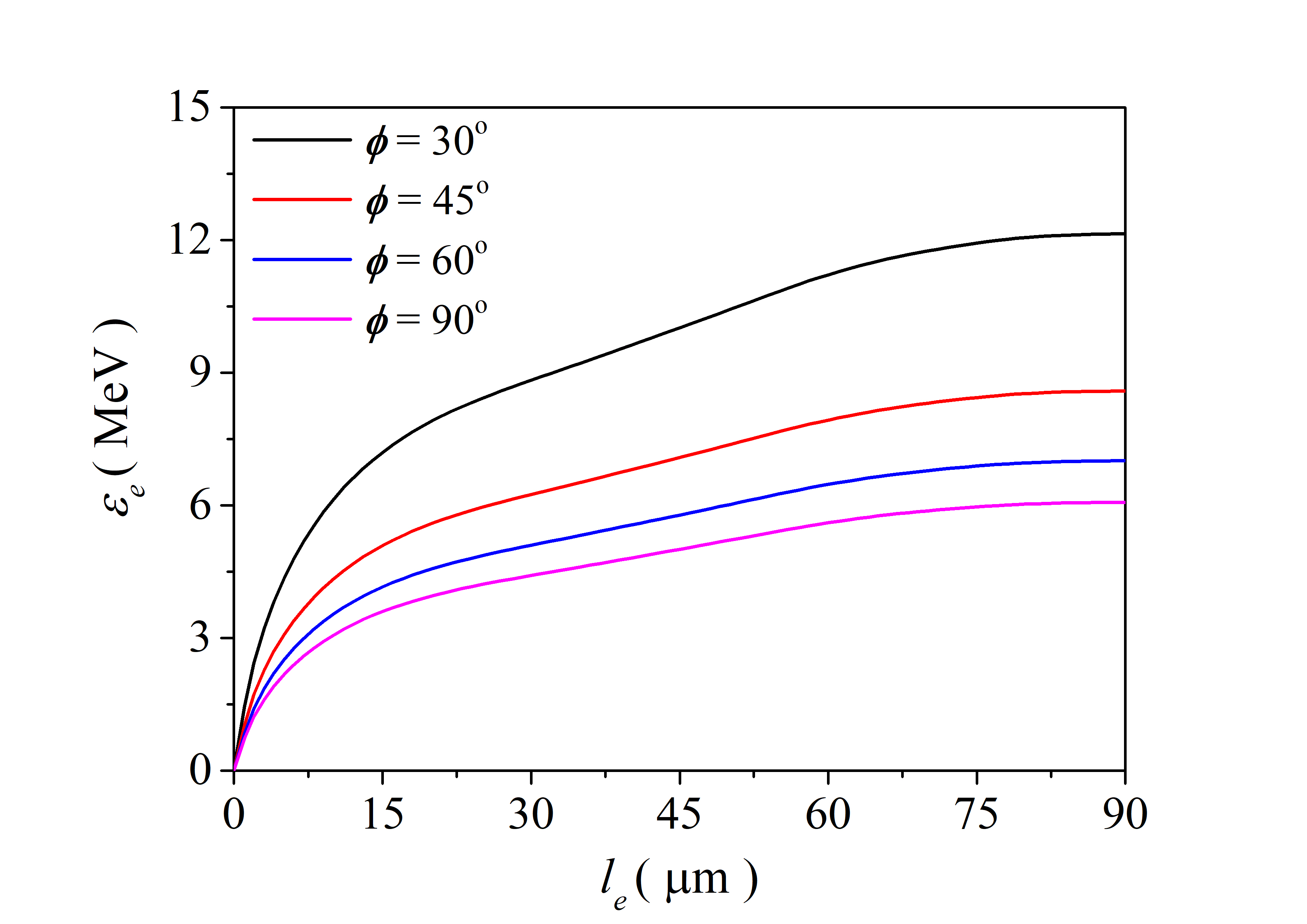}
    \caption{In the case of the electron penetrating the $E_s$ with different $\phi$ (the angle at which the electron enters the $E_s$), the relation between the electron threshold kinetic energy $\varepsilon_e$ and the transverse location $l_e$ where the electron could be pulled back into the plasma. }
    \label{fig2}
\end{figure}

The $E_s$, which propagates along the plasma surface \cite{Li2006} at a velocity close to the light speed, can guide the hot electrons in group B to oscillate transversely and propagate longitudinally as shown in Fig. 1d. Every time the electron passes through the transverse interfaces of the plasma, THz radiation can be emitted. This behaves like a wiggler in traditional light sources \cite{bilderback2005review}. Therefore, the plasma with transverse sheath fields $E_s$ can be considered as a wiggler for the electrons in group B. The reciprocating (wiggler) period of the electron can be expressed as
\begin{equation}
  \tau_r = \frac{4l_s + 2l_t}{v_t}. 
\label{equ3}
\end{equation} 
Here $l_s$ is the transverse distance between the plasma interface and the point where the electron turn around, $l_t$ is the thickness of the plasma, and $v_t$ is the median transverse velocity of the electrons. $l_s$ is determined by the transverse momentum of electrons and the strength of $E_s$. The radiation frequency $f$ is closely related to the wiggler frequency $f_w = 1/\tau_r$, then we have
\begin{equation}
  f = \frac{f_w (2\gamma^2)}{(1+K^2/2)}  =  \frac{v_t(2\gamma^2)}{(4l_s + 2l_t)(1+K^2/2)}, \label{equ4}
\end{equation} 
where $\gamma$ is the election Lorentz factor, $K$ is wiggle (undulator) parameter \cite{Jackson}. Normally, $l_s$ is several micro-meters \cite{mora2003}, while the changes on $\gamma$, $K$ and $v_t$ can be ignored for the same laser intensity and plasma density, then $l_t$ is the only parameter could be varied. Hence, one can regulate the frequency of the THz radiation by changing the plasma thickness $l_t$ according to Eq. (\ref{equ4}). Since the length of the electron beam is close to the laser pulse duration which is only 32 fs here, the radiation satisfies time coherence condition in the THz frequencies \cite{liao2019}.

\subsection{Electron kinetic}\label{subsec2}

To generate electrons in group B, one should use a plasma with limited transverse size to ensure the occurrence of  reciprocating motion. Simulations, with plasma of different lengths $l_p$ (fixed the plasma thickness $l_t$ to 30 $\mu$m), were performed to see the effect of plasma length on the electron dynamics. From the simulations with $l_p \ge $ 50 $\mu$m, it was found that more than $75\%$ of the laser energy was converted to the hot electrons (electron kinetic energy $e_k \ge $ 0.5 MeV). While in the case of $l_p$ = 10 $\mu$m, there were few electrons participating in reciprocating motion and most of the electrons could be classified to group A. 

%%%%%%%% Figure 3 %%%%%%%%

\begin{figure}[htb]
    \centering
    \includegraphics[width=\linewidth]{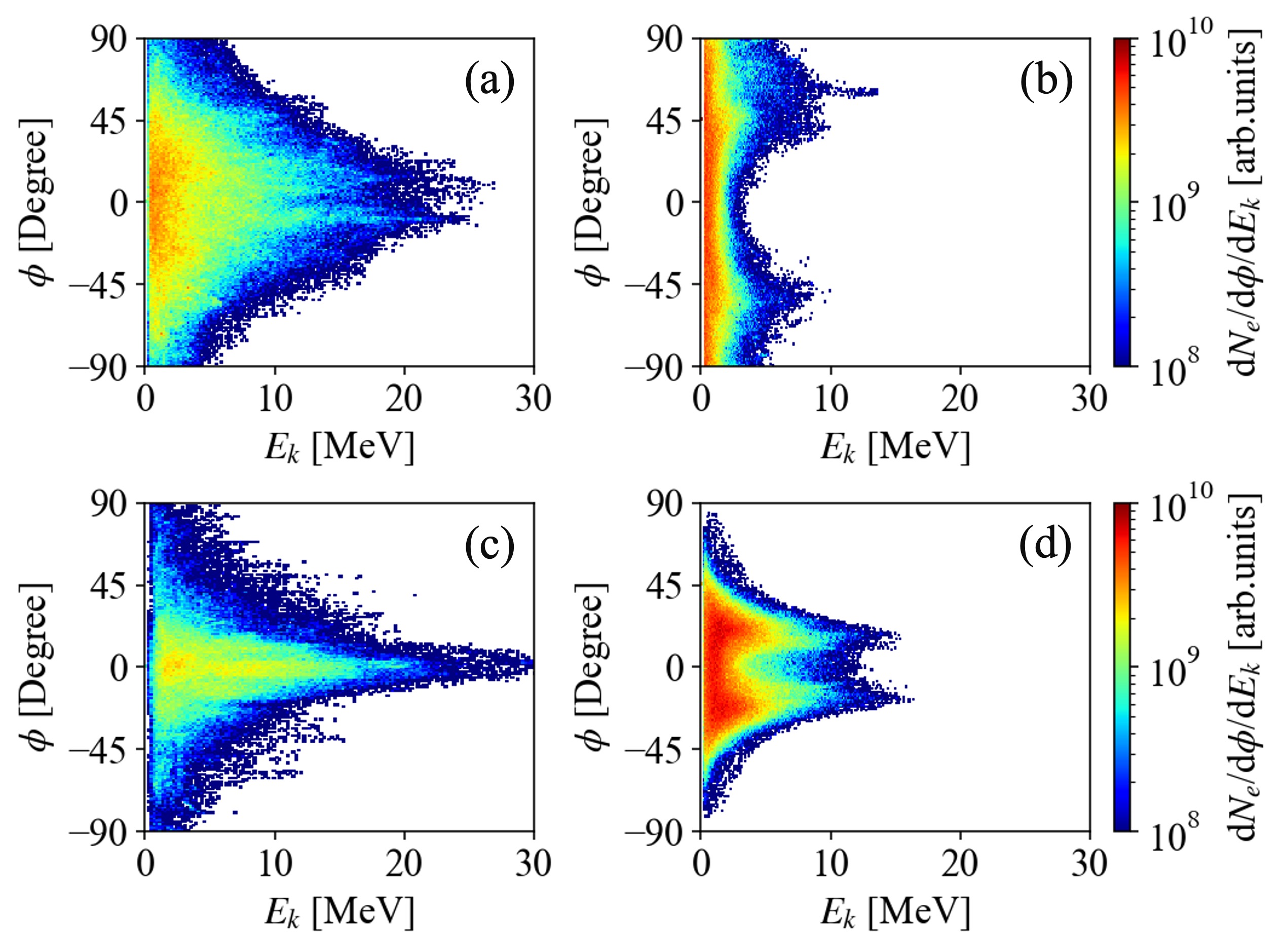}
    \caption{The angular-spectra distribution of the hot electrons. The electrons from a plasma length of $l_p$ =  50 $\mu m$ (a) collected by a screen with a radius of 55 $\mu m$  in the first 270 fs of the simulation could be classified to group A, (b) and the electrons behind group A can be assigned to group B. The electrons from a plasma length of $l_p$ =  200 $\mu m$ (c) in group A collected by a screen whose radius is 205 $\mu m$ in the first 770 fs, and (d) the electrons in group B.}
    \label{fig3}
\end{figure}

When $l_p$ was increased to 50 $\mu$m, one could easily distinguish the electrons in group B from group A, as plotted in Figs. 3a and 3b. Figure 3a shows a typical angular-spectra distribution of the electrons, with a large divergence angle, accelerated by the laser ponderomotive force \cite{wilks1992}. This figure is very similar to the result of $l_p$ = 10 $\mu$m. Under the action of $E_s$, the reciprocating motion of the electrons is facilitated and obvious changes on the angular-spectra distribution can be seen from Fig. 3b. In this case, the electron beam had a divergence angle of $\sim60^\circ$, which indicates that most of the electrons penetrate the transverse sheath fields with an angle of $\phi \approx 60^\circ$. Hence, the threshold kinetic energy of the electron $\varepsilon_e$ is about 8 MeV, which is consistent with the theoretical value predicted by Eq. (\ref{equ2}) as shown in Fig. 2. If $l_p$ was further increased, such as $l_p$ = 200 $\mu$m, more electrons of relatively low energy in group A come into group B as shown in Fig. 3c, and the number of electrons in group B increases with $l_p$. As a result, the electron number shown in Fig. 3d is larger than that in Fig. 3b. It is noted that the electrons in group B were manipulated and collimated into a smaller divergence angle by $E_s$, and the electron threshold kinetic energy $\varepsilon_e$ could be much higher as predicted by Eq. (\ref{equ2}). 

We used a semicircular receiving screen, whose radius was 250 $\mu$m with center locating at the midpoint of the plasma right interface, to collect the radiation field. Then, the angular-spectra distribution of the THz source could be obtained from the radiation field through Fourier transform \cite{cai2022}. In this work, the angular spectrum method \cite{lalor1968,ritter2014,cai2022} was employed to remove the near-field radiation, and then all the results of the THz source could be recognized as far-field radiation. 

%%%%%%%% Figure 4 %%%%%%%%

\begin{figure}[htb]
    \centering
    \includegraphics[width=\linewidth]{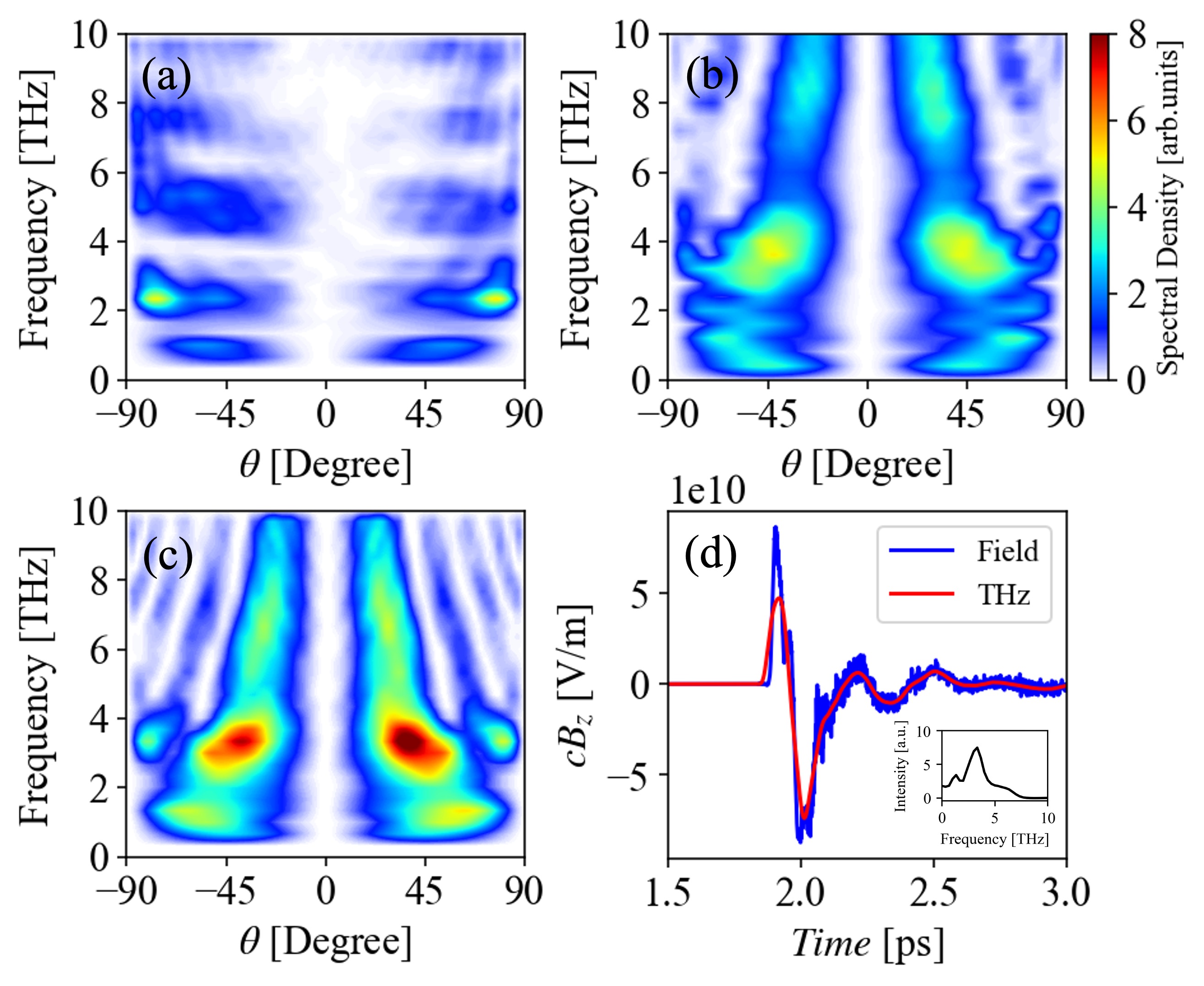}
    \caption{Simulation results from the plasma of different lengths $l_p$, while the thickness $l_t$ was fixed to 30 $\mu m$. The angular-spectra distribution of the THz pulses from (a) $l_p$=50 $\mu m$, (b) $l_p$=150 $\mu m$ and (c) $l_p$=300 $\mu m$. (d) The radiation field before filtering (blue line) and the field of the THz pulse (red line) collected at $37^\circ$ from the simulation of $l_p$=300 $\mu m$.}
    \label{fig4}
\end{figure}

In the simulation with $l_p$ = 150 $\mu$m, the median longitudinal (transverse) velocity $v_l$ ($v_t$) of the electrons in group B was 0.94c (0.85c). The transverse size of $E_s$ was about 2 $\mu$m (at FWHM), and therefore we could assume $l_s \approx$ 2 $\mu$m. According to Eq. (\ref{equ3}), one can get $\tau_r$ = 270 fs and $\lambda_w  \approx \tau_r \times v_l$ = 79.4 $\mu$m. The wiggler period $\tau_w = \tau_r$ corresponds to a wiggle frequency $f_w$= 3.7 THz, and the Lorentz factor $\gamma$ for the longitudinal velocity $v_l$ was 2.93. By tracking the electromagnetic fields exerting on the electrons in group B, we got an average value $K\approx 6.0$ which decreased to 5.5 when $l_s$ increased to 80 $\mu$m. According to Eq. (\ref{equ4}), the radiation frequency was $f$ = 3.34 THz, which is also the same as the center-frequency 3.35 THz from the simulation. Hence, the theoretical model can accurately predict the terahertz frequency obtained from the simulations. 

\subsection{Spectrum of terahertz}\label{subsec3}

Figures 4a-c show the angular-spectra distribution of the THz pulses from the plasma lengths $l_p$ = 50, 150 and 300 $\mu$m. In the condition of $l_p$ shorter than $\lambda_w $, the role of electrons in group B could be ignored and the generation of THz source is dominated by target rear CTR \cite{liao2015,liao2019,liao2020towards} from the electrons in group A. Hence, the THz source had a large divergence angle (pointed at 88$^\circ$) and a small energy conversion efficiency of 0.45$\%$ as shown in Fig. 4a.

Every time the electron passes through the transverse interfaces of the plasma, THz radiation can be emitted and become stronger in the case of the electrons in group B oscillating for several periods. When $l_p$ was enlarged to 150 $\mu$m, more than $1.3\%$ of the laser energy was converted to the THz radiation and the characteristics of the narrow spectrum became obvious as shown in Fig. 4b. The simulation results indicated that $l_p \ge$ 150 $\mu$m was more conducive to obtain THz pulses of narrow spectrum as shown in Figs. 4b and 4c. As the electrons in group B are guided to oscillate along the plasma by $E_s$, the resulting THz pulse could be better collimated than the other intense laser-plasma based THz sources \cite{liao2015,yi2019,liao2020towards}. It was found that the THz pulse which pointed at $\theta = 37^\circ$ could be collimated into a divergence angle of $\sim 20^\circ$ as shown in Fig. 4c. Since $\theta > 1/\gamma$, we call it wiggler instead of undulator. Then we collected the radiation field at $37^\circ$ from the simulation of $l_p$ = 300 $\mu$m, and filtered the THz signal (frequency ranged 0.1-10 THz) from the radiation field as shown in Fig. 4d. The THz field was higher than 80 GV/m, which was much greater than those from THz sources driven by similar laser parameters \cite{yi2019,liao2020towards}, after the THz pulse propagating 250 $\mu$m off the plasma. Owing to the THz radiation comes from the wiggler process, the THz electric field waveform of multi-cycles \cite{tian2017} was completely different from the half-cycle THz sources driven by intense laser and plasma \cite{liao2015,yi2019,liao2020towards}.

%%%%%%%% Figure 5 %%%%%%%%

\begin{figure}[htb]
    \centering
    \includegraphics[width=\linewidth]{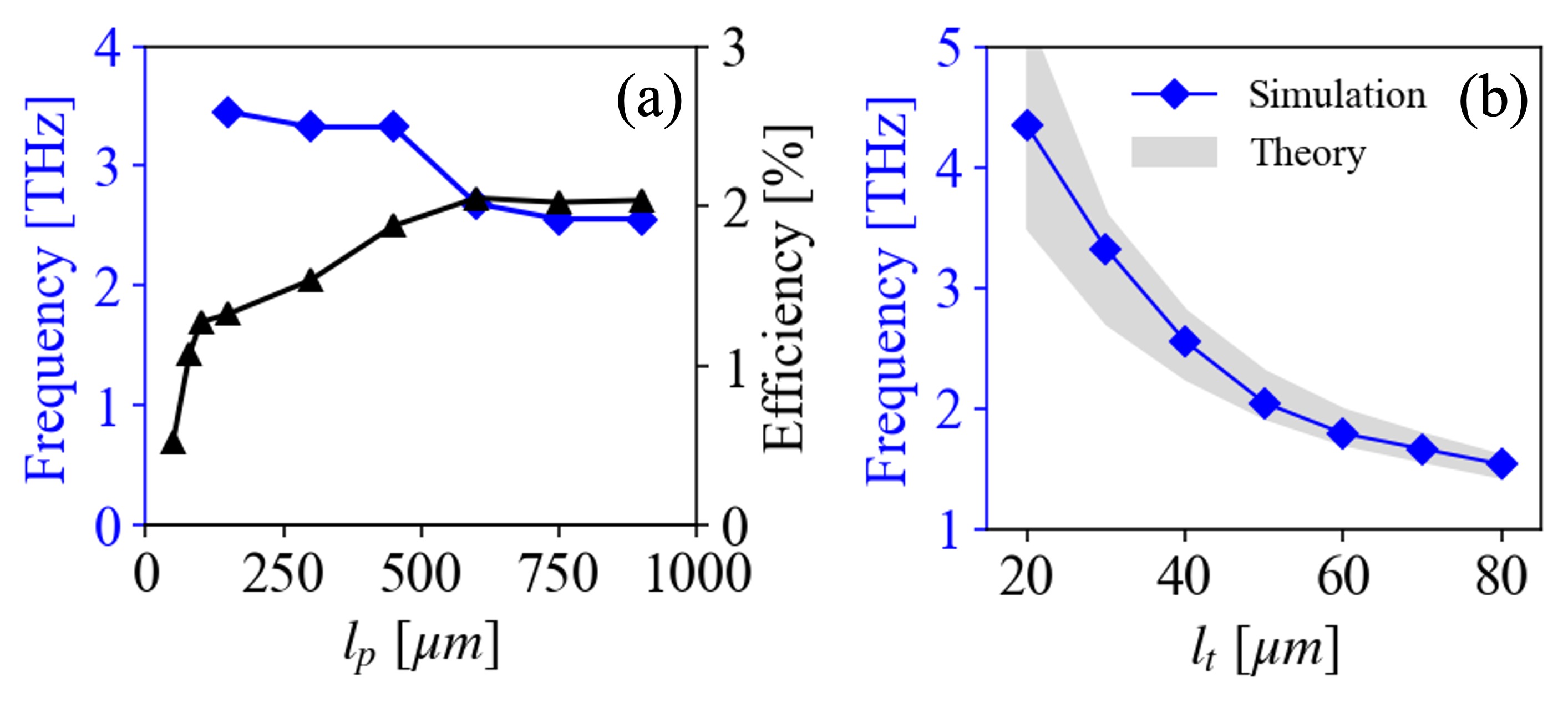}
    \caption{(a) The center-frequency of the THz source and the laser-THz energy conversion efficiency from the simulations with plasmas of different lengths $l_p$ from 50 to 900 $\mu m$, while $l_t$ was fixed to 30 $\mu m$. (b) The center-frequency of the THz source from the simulations (blue dotted line) and the theoretical model of Eq. (\ref{equ4}) for the plasma thickness $l_t$ changing from 20 to 80 $\mu m$ (light black shadow).}
    \label{fig5}
\end{figure}

\subsection{Tunable of frequency}\label{subsec4}

From Figs. 3a and 3c, one can see that the electrons in group A had a large divergence angle. The electron density $n_e$ decreases with the increase of transmission distance, and $E_s$ becomes weaker for the longer plasma according to Eq. (\ref{equ1}), in which case the electrons in group B are more difficult to be pulled back to the plasma. As a result of enhancing $l_p$, $l_s$ becomes larger and more electrons no longer participates in the wiggler motion. Hence, the center-frequency decreases and the growth of energy conversion efficiency gradually slows down with the enhancement of $l_p$. Simulations by varying $l_p$ from 50 to 900 $\mu$m have demonstrated the effect on the center-frequency as shown in Fig. 5a. From the figure, one can see that the center-frequency of the THz pulse was tunable from 3.5 to 2.6 THz by changing the plasma length $l_p$ from 150 to 900 $\mu$m. On the other hand, the laser-THz energy conversion efficiency, which increased with $l_p$ and became saturated after 600 $\mu$m as shown in  Fig. 5a, was 2.05$\%$. The saturation length may be longer than 600 $\mu$m since a small portion of THz source came out of the simulation window, in the case of longer $l_p$, due to the limitation of computational ability. 

Equation (\ref{equ4}) indicates that we could regulate the center-frequency of the THz pulse by changing the thickness of the plasma $l_t$. More simulations were performed by changing $l_t$ from 20 to 80 $\mu$m (fixed $l_p$ to 300 $\mu$m). It was found that the laser-THz energy conversion efficiency was almost the same for different $l_t$, while the center frequency of the THz source decreased with the increase of $l_t$ as shown in Fig. 5b. For different $l_t$, $v_t$ can be obtained from the corresponding simulations, and the simulation results also shown $l_s$ ranging 0.5 to 5 $\mu$m. We assumed that $K$ decreased with the increase of $l_s$ from 6.0 to 5.5 for the above cases, then calculated the center frequency of the THz pulse according to Eq. (\ref{equ4}) and plotted them in Fig. 5b. From the figure, one can see that the theoretical model developed here works very well in predicting the generation of center-frequency tunable THz pulse with a laser driven plasma wiggler.

\section{Discussion}\label{sec12}

To validate the physical results, we conducted 3D Particle-in-Cell (PIC) simulations using a box size of $1920 \times 800 \times 800$ and a resolution of $0.15 \lambda \times 0.25 \lambda \times 0.25 \lambda$ (further details can be found in the supplementary materials). The 3D simulation produced a spectrum in the cross-section that closely resembles the one obtained from the 2D simulation. Additionally, the spatial distribution exhibits a quasi-2D pattern, indicating similarities between the two simulations.

Parametric simulations were also carried out to see the effects of laser intensity, plasma density and density gradient at the boundary (more details at supplementary materials). It was found that the energy conversion efficiency almost unchanged for the plasma of the same thickness under a relativistic intense laser ($a_0 \ge 1.0$). While the center frequency, which decreased from 3.35 to 1.7 THz, would further reduce under smaller $a_0$ due to the drop of electron $v_t$. Since a low density plasma is more beneficial to improving the energy efficiency from laser pulse to relativistic electrons, the laser-THz energy conversion efficiency was higher in the case of lower density plasma. However, the center frequency of the THz source was almost the same. In reality, the plasma has a density gradient at the boundaries. Simulations using a pre-expand plasma with a scale length varied from 0.1 to 1 $\mu$m showed that there was almost no changes on the THz angular-spectra distribution. The parametric scans demonstrated the validity of this novel laser-driven plasma wiggler for the generation of high-power and center-frequency tunable THz pulses over a wide range of laser and plasma parameters. With this method, one could produce ultra-high power THz pulse of $\sim$ 200 mJ by using a laser pulse of $\sim 10$ J. Such a state-of-the-art THz source would initiate future applications of high-power THz pulse basing on a compact laser.

\section{Conclusion}\label{sec13}

In summary, we propose a laser-driven plasma wiggler for efficiently generating a high-power, collimated, narrow-band and center-frequency tunable THz pulse, by manipulating electrons reciprocating motion. A theoretical model is developed to describe the physical principle of realizing the center-frequency tunable THz pulse. According to the model and PIC simulations, the center frequency of the THz pulse corresponds strictly to the reciprocating motion period of the electron beam. Simulations indicate that the center frequency of the THz pulses can be tuned from 4.4 to 1.5 THz as the plasma thickness changed from 20 to 80 $\mu$m. Meanwhile, the THz pulse is collimated into a divergence angle of $\sim20^\circ$, enabling a laser-THz conversion efficiency $ > 2.0 \%$ and an ultra-strong field strength of over 80 GV/m, driven by a table-top laser of $\sim430$ mJ. This method could address a long-standing challenge in THz science and generate a state-of-the-art THz source over a wide range of laser and plasma parameters.

\appendix
%\appendices

\section{Particle-in-cell simulation setup}

Numerical simulations were performed by using the two-dimensional PIC codes Epoch \cite{arber2015} and Smilei \cite{derouillat2018} to verify this scheme. The simulation window X $\times$ Y = 600 $\mu$m $\times$ 570 $\mu$m was divided into $12500\times 7125$ cells. A laser pulse, with Gaussian spatial and $sin^2$ temporal profiles, wavelength $\lambda _0$ = 800 nm, waist $w_0$ = 7.2 $\mu$m, normalized intensity $a_0$ = 3, polarized in Y direction and duration of 32 fs at FWHM (full width at half maximum), was used as the driving source. Such a laser pulse can be produced by a compact laser \cite{danson2019}. We used a block-shaped near-critical density plasma (electron density $n_e = 0.2n_c$, where $n_c$ is the critical density) whose length and thickness were adjustable. Such a plasma could be made of carbon nanotube foams \cite{ma2007,shou2023}. 24 (4) macro-electrons (C$^{6+}$) were initialized into each cell. The simulation time was varied with plasma length $l_p$ to make sure that we had more than 1 picosecond to collect the THz signal, which should be enough to cover the whole interaction process for the generation of THz sources. Meanwhile, the zero padding method \cite{harris2020} was used to further improve the resolution of the THz signal. 

\section*{Acknowledgments}

This work was supported by the Natural Science Foundation of China (Grant Nos. 11921006, 12175058), Beijing distinguished young scientist program and National Grand Instrument Project No. SQ2019YFF01014400. The PIC code EPOCH was in part funded by United Kingdom EPSRC Grant Nos. EP/G054950/1, EP/G056803/1, EP/G055165/1, and EP/M022463/1. We thanks the helpful discussions with Prof. Z. Najmudin from Imperial College London and Prof. Wenjun Ma from Peking University.

%\bibliographystyle{unsrt}
%\bibliography{cite}

\end{document}